\documentclass[referee]{raa}            

\usepackage{graphicx,times}             
\usepackage{natbib}
\usepackage{amssymb,amsmath}
\bibpunct{(}{)}{;}{a}{}{,}

\usepackage[pagebackref=true]{hyperref}
\hypersetup{colorlinks = true, linkcolor = green, anchorcolor = red, citecolor = blue, filecolor = red, urlcolor = red}

\begin{document}

   \title{Early-fusion Based Pulsar Identification with Smart Under-sampling
}

   \volnopage{Vol.0 (20xx) No.0, 000--000}      
   \setcounter{page}{1}          

   \author{Shi-Chuan Zhang
      \inst{1,2}
     \and Xiang-Cong Kong
      \inst{1,2}
   \and Yue-Ying Zhou
      \inst{1}
   \and Ling-Yao Chen
      \inst{1}
   \and Xiao-Ying Zheng
      \inst{1,2}
   \and Chun-Ling Xu
      \inst{1}
   \and Bao-Qiang Lao
      \inst{3}
   \and Tao An
      \inst{3}
   }

   \institute{Shanghai Advanced Research Institute, Chinese Academy of Sciences,
             Shanghai 201210, China; {\it Corresponding author: X. Zheng (zhengxy@sari.ac.cn).}\\
        \and
            University of Chinese Academy of Sciences\\
        \and
             Shanghai Astronomical Observatory, Chinese Academy of Sciences\\
\vs\no
   {\small Received~~; accepted~~}}

\abstract{ The discovery of pulsars is of great significance in the field of physics and astronomy. As the astronomical equipment produces a large amount of pulsar data, an algorithm for automatically identifying pulsars becomes urgent. We propose a deep learning framework for pulsar recognition.
In response to the extreme imbalance between positive and negative examples and the hard negative sample issue presented in the HTRU Medlat Training Data, there are two coping strategies in our framework: the smart under-sampling and the improved loss function. We also apply the early-fusion strategy to integrate features obtained from different attributes before classification to improve the performance. To our best knowledge, this is the first study that integrates these strategies and techniques together in pulsar recognition. The experiment results show that our framework outperforms previous works with
the respect to either the training time or $F1$ score.
We can not only speed up the training time by $10X$ compared with the state-of-the-art work, but also get a competitive result in terms of $F1$ score.
\keywords{methods: data analysis — (stars:)pulsars: general — techniques: image processing }}

    \authorrunning{S.-C. Zhang, Y.-C. Kong, Y.-Y. Zhou et al.}       
    \titlerunning{Early Fusion Based Pulsar Identification with Smart Under-sampling}  

    \maketitle

\section{Introduction}
\label{sec:introduction}

Pulsar is a high-speed spinning neutron star which can continuously emit electromagnetic pulse signals. In physical and astronomical studies,
the cosmic clocks produced by pulsars can be used as a galactic-scale detector
for many fundamental physics applications, including gravitational wave.
There are several modern radio telescopes survey projects, such as Parkes Multi-beam Pulsar Survey (PMPS) \cite{keane2010further}, High Time Resolution Universe (HTRU) \cite{keith2010high}, Pulsar Arecibo L-band Feed Array survey (PALFA) \cite{cordes2006arecibo} that are actively scanning the sky and collecting signals for potential pulsar candidates.
In a pulsar search procedure, periodic broadband
signals exhibiting signs of dispersion from the universe space
are collected by the modern radio telescope surveys.
These signals are processed and recorded as a pulsar
candidate collection of diagnostic plots and summary statistics if
they meet some criteria.
The pulsar candidate collection is further analyzed manually or
automatically to select the prospective candidates for further verification.
A pulsar candidate contains a set of diagnostic values and graphical representations,
including the time-versus-phase (TPP), the
frequency-versus-phase (FPP), the summed profile plot (SPP), and the dispersion measure (DM) curve etc.
Due to the improving survey specifications, there is an exponential rise in the pulsar candidate numbers and data volumes.
The machine learning techniques have been introduced recently to mitigate the scalability issue of the pulsar identification problem \cite{Yao2017Pulsar} \cite{Zhang2019Pulsar} \cite{Wang2017Weighting} \cite{Morello2014SPINN} \cite{osti_22348078} \cite{Tao2019Science} \cite{Bethapudi_2018}.

\begin{figure}[t!]
\centering{
\includegraphics[width=9.5cm]{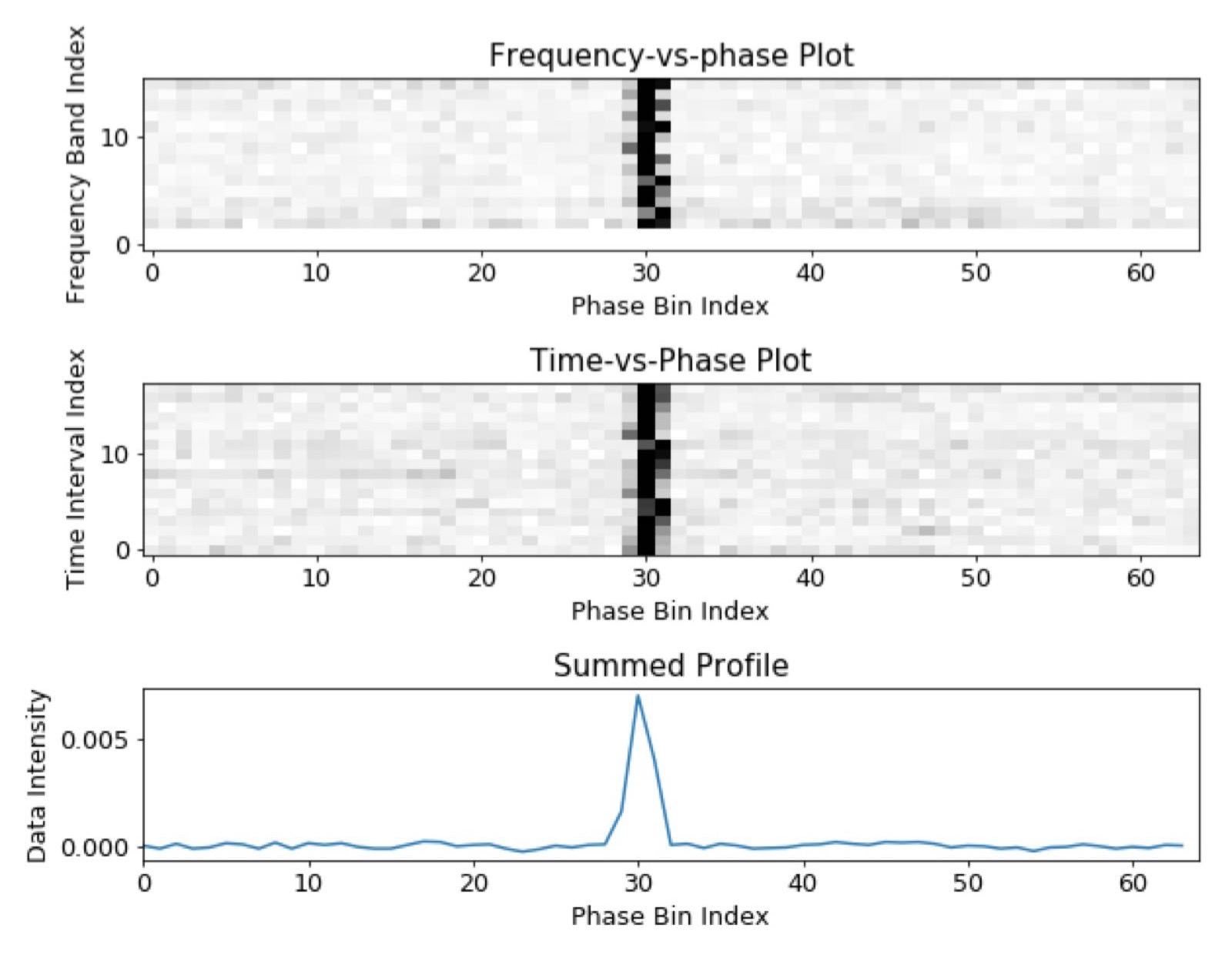}}
\caption{A typical positive pulsar candidate in the HTRU Medlat dataset. The periodical signals are processed by the pipeline software to produce a set of diagnostic values and graphical representations including FPP, TPP and SPP. For a real pulsar, there is a bold vertical stripe in both FPP and TPP, and a significant peak in SPP.
\label{fig:positive}}
\end{figure}

\begin{figure}[t!]
\centering{
\includegraphics[width=9.5cm]{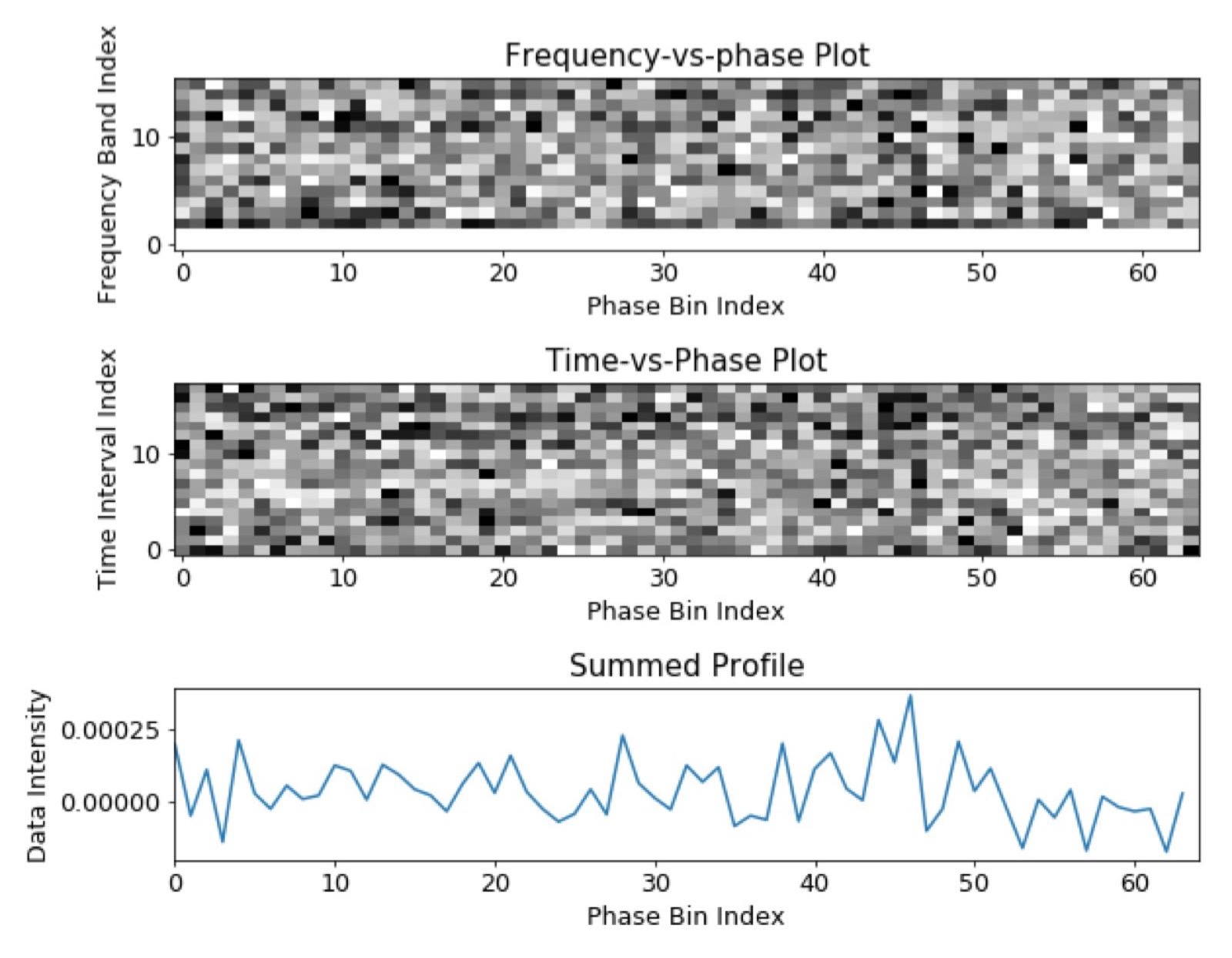}}
\caption{An typical negative candidate in the HTRU Medlat dataset. For an RFI signal, there is no bold vertical stripes in neither FPP nor TPP, or no significant peaks in SPP.
\label{fig:negative}}
\end{figure}

\begin{figure}[!t]
\centering{
\includegraphics[width=8.5cm]{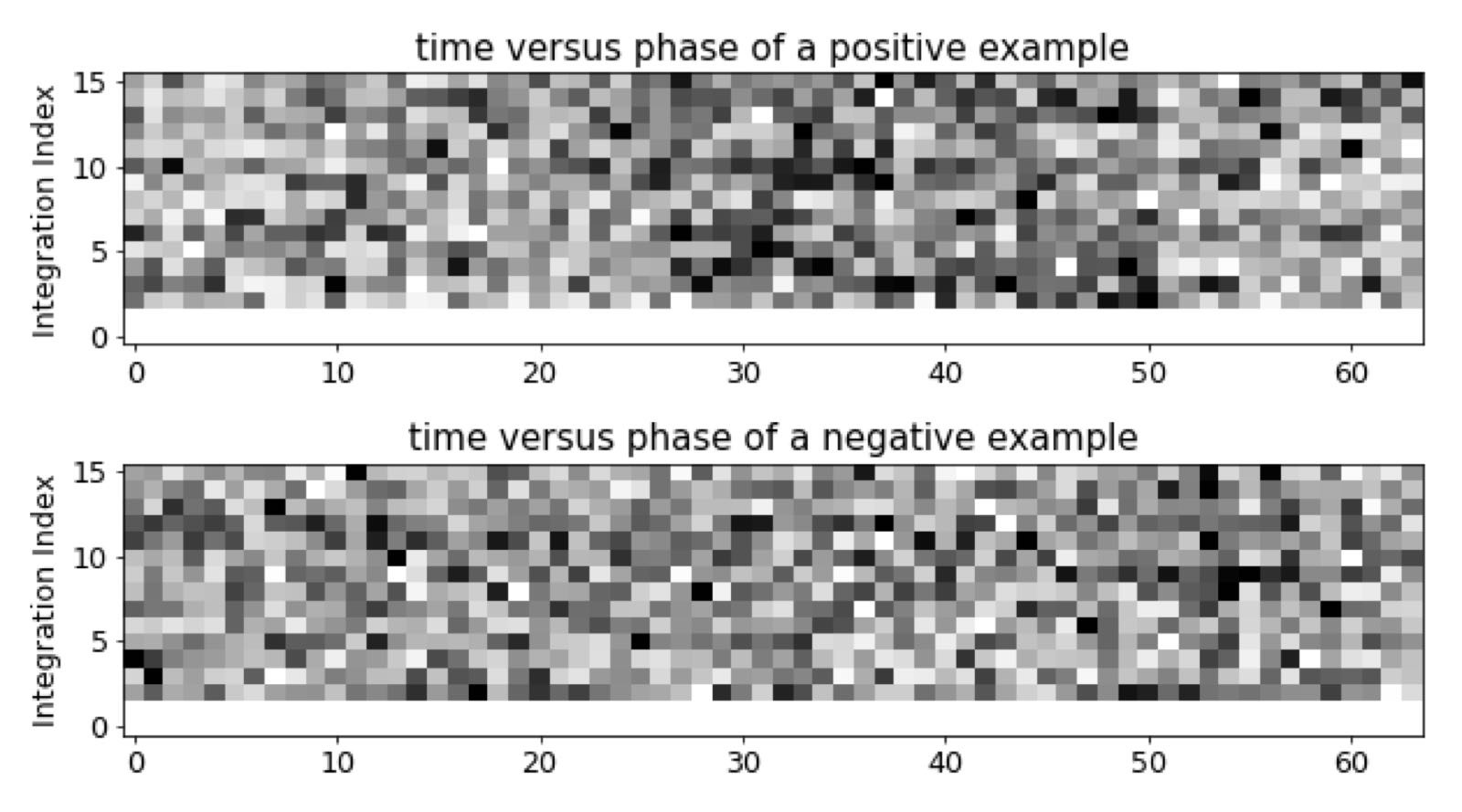}}
\caption{Hard samples. The top picture is a TPP of a positive pulsar candidate; and
the bottom picture is a TPP of a negative pulsar candidate. There is no obvious bold vertical stripes in the TPP of the positive sample. Both the two TPPs are very similar and are hard to identify.
\label{fig:hard_sample}}
\end{figure}

There are some open-source pulsar candidate sets such as HTRU Medlat Training Data collected by
the HTRU survey \cite{Morello2014SPINN} and FAST Label Data collected by the Commensal Radio Astronomy Five-hundred-meter Aperture Spherical radio Telescope (FAST) Survey \cite{Wang2019}.
The HTRU Medlat dataset is a collection of labeled pulsar candidates from the intermediate galactic latitude part of the HTRU survey, which contains precisely 1,196 known pulsar candidates from 521 distinct sources, and 89,996 non-pulsar candidates.
The FAST dataset contains $837$ known pulsar candidates and $998$ non-pulsar candidates in the training set;
and $326$ pulsar candidates and $13321$ non-pulsar candidates in the testing set, respectively.
Hence, the skewness of the dataset is not very severe compared with HTRU.
In this paper, we focus on the open-source pulsar candidate set of HTRU Medlat Training Data.
We use the code provided by \cite{Morello2014SPINN} to generate the attributes of FPP, TPP and SPP for each
pulsar candidate, and show two sample candidates in Fig. \ref{fig:positive} and Fig. \ref{fig:negative}.
In Fig. \ref{fig:positive}, in both FPP and TPP, there is a bold vertical stripe; and there is also a significant peak in SPP. Thus this pulsar candidate can be labeled as a positive candidate and
will be selected for further verification by astronomers.
In Fig. \ref{fig:negative}, there is no bold vertical stripe in neither FPP nor TPP, or no significant peak in SPP. Thus this pulsar candidate can be labeled as negative and will be dropped.
As positive and negative candidates present significant differences in FPP, TPP and SPP, a machine learning based pulsar candidate selection approaches can be used to screen the candidates
automatically and will be investigated in our paper.

Automatic recognition is an efficient approach to improve pulsar filter efficiency.
In general, deep learning is robust with a wide spectrum of applications based on the huge amounts of data, the fast-computing devices and the large capacity storage.
However, sometimes it does not work well when the numbers of samples are not balanced between different classes. This phenomenon is called a class imbalance problem.
It is claimed that the classification performance of
the same deep learning model can be substantially
different with the balanced and imbalanced data sets \cite{Padma2011Performance}.
With the imbalanced data, the model has preference for the class with more samples than other classes. Therefore, the accuracy of test tends to be high, but the recall of the minor class with less samples is low. However, the recall of minor class is usually an important indicator of performance.
Unfortunately, the pulsar identification problem suffers from the class imbalance.
In a pulsar candidate set, most of the samples are radio frequency interference (RFI) signals that are noise, i.e., negative samples; and only a small set of samples is real pulsars, i.e., positive
samples. The ratio of positive sample to negative samples can be up to $1:90$. Therefore, the negative samples dominate the pulsar candidate dataset and become a huge challenge for pulsar recognition.
Furthermore, even for positive pulsars, there exist hard samples. As shown in Fig. \ref{fig:positive} and Fig. \ref{fig:negative}, there are some real pulsars that present
significant difference in diagnostic plots and are easy to identify; and there also exist
some hard negative noises that are very similar to real pulsars and are very hard to classify as
shown in Fig. \ref{fig:hard_sample}.
In Fig. \ref{fig:hard_sample}, the top picture shows the time-versus-phase (TPP) plot of a positive pulsar
candidate; and the bottom picture shows the TPP of a negative candidate.
Since there is no obvious bold vertical stripe in the TPP of the positive sample, the two TPPs are very similar and hard to differentiate between each other. These candidates are called hard samples.

To overcome the above challenges, we propose a deep-learning based framework
for the binary-classification pulsar recognition with imbalanced classes.
The framework consists of three stages.
At Stage 1, we apply a \textit{smart under-sampling} method which was first proposed by \cite{vannucci2016smart}.
The smart under-sampling deletes about $86\%$ of the negative samples
and retains almost all positive samples by calculating the Mahalanobis distance for each sample.
With this smart under-sampling, we can reduce the ratio of positive to negative samples from $1:90$
to $1:10$, which greatly mitigates the class imbalance.
At Stage 2, we apply deep learning network models to extract features from the dataset.
We select three attributes from the $17$ attributes of pulsars, i.e., TPP, FPP and SPP, and extract
features for each of the attributes.
We apply the $\alpha$-balanced loss function to further help the class imbalance issue, and
propose to use a $\gamma$ parameter to down-weight the
easy examples and focus on the hard negative examples in the loss function.
At Stage 3, we propose the simple attention mechanism to learn the weights of the three features obtained at Stage 2, and blend the features to form a single feature.
This is called the \textit{early-fusion} strategy, which was first proposed in \cite{Gunes2006Affect}.
In the early-fusion, the features are blended before classification; in the meanwhile, the late-fusion strategy proposed in \cite{osti_22348078} and \cite{li2018hierarchical} blends the classification results after the classifying operation.
Finally, the blended feature is fed into a binary classifier to filter the positive candidates

We conclude our contribution. We propose a smart under-sampling to solve the class
imbalance. We apply the early-fusion based simple attention mechanism to integrate different features obtained from three
attributes of pulsars for further classification. We improve the loss function to
further help the class imbalance and the hard sample issues.
The rest of the paper is organized as follows. In section \ref{sec:related}, we investigate the related works.
We present our deep learning based pulsar identification framework in Section \ref{sec:method}.
In Section \ref{sec:experiment}, our framework is evaluated by comparing with
the state-of-the-art works. We draw a conclusion in Section \ref{sec:conclusion}.

\subsection{Related Works}
\label{sec:related}

There are some successful works for pulsar recognition with machine learning. The earlier approaches focused on the statistical features manually designed by astronomers \cite{Yao2017Pulsar} \cite{Wang2017Weighting}. Recently, some data-driven deep learning models have emerged.
Zhu et al. proposed a pulsar image-based
classification system (PICS) \cite{osti_22348078}.
PICS uses the single-hidden-layer artificial neural network
(ANN), the support vector machines (SVMs), and the convolutional neural networks (CNNs) to process different
features, and assembles the networks together with a logistic
regression classifier to construct the PICS.
The overall architecture does not consider the imbalanced issue. The plain CNN networks are used to
classify the TPP and FPP plots, and a simple weighted strategy is used to assemble different classifiers.
Nevertheless, the work proposed by Zhu et al. demonstrates the capability and
advantages of the CNNs for pulsar candidate selection task.
In \cite{Wang2019}, Wang et al. further improved PICS by replacing the CNNs with a residual network model
comprising $15$ layers. The work in \cite{Wang2019} does not consider the class imbalance either.
In \cite{Guo2017Pulsar}, Guo et al. used the deep convolution generative adversarial networks (DCGAN) and the support vector machine (SVM) to extract features and classify pulsar candidates. Li et al. used a hierarchical model to identify the pulsars with four diagnostic plots as inputs in \cite{li2018hierarchical}. They reduced the running time of the complex model with a pseudo-inverse learning algorithm. There are some other data-driven methods for pulsar recognition such as \cite{osti_22348078} and \cite{Morello2014SPINN}. The advantage of data-driven approaches is that they do not require to manually extract features and finish tasks. The framework proposed in this paper is a data-driven approach too.

A lot of works have been investigated to cope with the unbalanced classification. In the survey of
the unbalanced binary classification \cite{gao2018study}, Gao et al. grouped different approaches into two categories: data-level methods and algorithm-level methods. The data-level methods include re-sampling and
ensemble learning. The re-sampling approach balances the data set by repeatedly sampling
positive samples \cite{Chawla2002} \cite{Han2005} \cite{He2008},
or under-sampling negative samples \cite{Wilson1972} \cite{IEEE1976}.
For example, Zhang et al. used the random over-sampling to classify pulsar candidates in \cite{Zhang2019Pulsar}.
Re-sampling algorithms with preference for specific samples are proposed in \cite{vannucci2016smart}, \cite{vannucci2017genetic} and \cite{vannucci2018self}.
The ensemble learning approach builds multiple subsets of the majority
class by random under-sampling. Each of the under-sampled majority subsets is united with the minority class to create a balanced data set.
Based on these balanced under-sampled data set, weak classifiers are trained and integrated through voting or weighting to build the overall classifier.
The work in \cite{Liu2009} proposes two ensemble learning models as EasyEnsemble and BalanceCascade.
EasyEnsemble combines the learners that are trained from multiple subsets of the majority class.
BalanceCascade trains the learners sequentially, where the majority class examples that are correctly classified by the current trained learners are removed at each training step.
Algorithm-level methods generally apply cost-sensitive functions to tune sample weights. In \cite{Zhou2017An} and \cite{Lin2017Focal}, the improved loss function is used for the single-target tracking problem and the objects detection, where the quantities of positive samples and negative samples are unbalanced.
A hybrid feature selection algorithm is proposed in \cite{Zhang2018} to process the unbalanced data problem when the data has multiple features.
The work in \cite{Guo2017Pulsar} uses the DCGAN architecture to generate pulsar samples to solve the unbalanced issue.

Next, we investigate feature fusion.
A proper feature fusion strategy can greatly improve the classification accuracy.  In \cite{Chaib2017Deep}, a discriminant correlation analysis (DCA) was adopted as the feature fusion strategy to refine the original features, which is more efficient than the traditional feature fusion strategies.
The attention mechanism is widely used in visual tasks.
A feature fusion method based attention mechanism was proposed for the video classification in \cite{long2018multimodal}.  In \cite{Wang2017Residual}, the author designed a residual attention network (RAN) to generate attention-aware features.

\section{A deep learning framework for pulsar identification}
\label{sec:method}

\begin{figure*}[!t]
\centering{
\includegraphics[width=\textwidth]{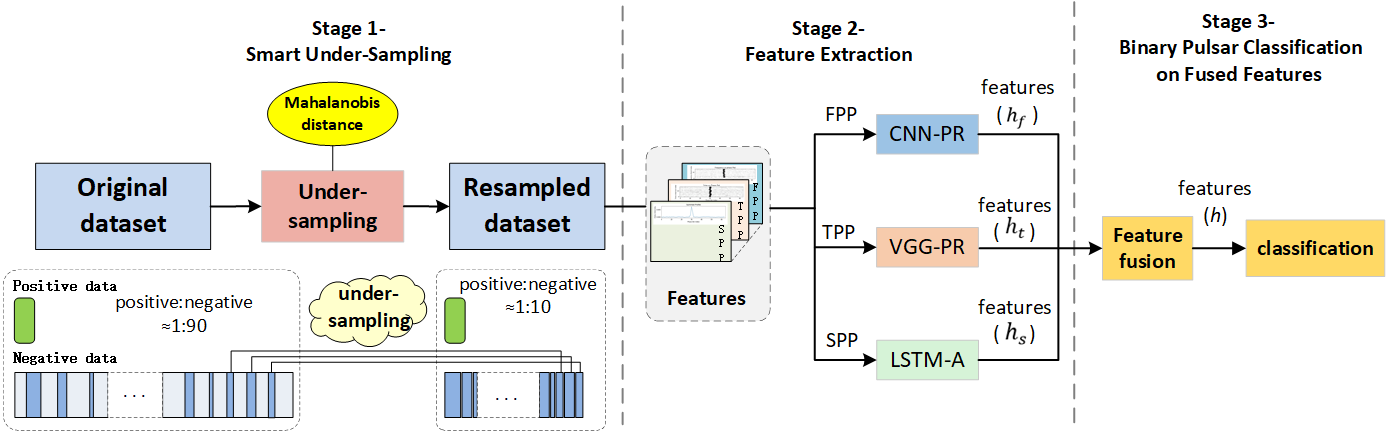}}
\caption{
The learning framework consists of three stages. At Stage 1, the original HTRU Madlat dataset is cleaned by the smart under-sampling. After cleaning, 85\% of data are washed away and the dropped data are almost negative samples. At Stage 2, the data after cleaning are used to train and test fine classifiers to extract features of diagnostic plots. At Stage 3, we train and test a neural network that can fuse features for classification. The fused features are finally fed into a binary pulsar/noise classifier.
\label{fig:framework}}
\end{figure*}

The challenge faced by pulsar candidate selection is the extreme imbalance
between positive and negatives samples in the dataset.
In this section, we propose a novel learning framework integrating three techniques including smart undersampling, improved cross entropy loss and attention-based feature fusion, to cope with the data imbalance as shown in Fig. \ref{fig:framework}.

\subsection{Stage 1: Smart Under-Sampling}
\label{sec:stage1}

The HTRU Medlat Training Data set contains 1,196 known pulsar samples and 89,996 non-pulsar samples, which results in a positive-to-negative-sample ratio of nearly $1:90$.
We propose to remove most of the non-pulsar samples from the dataset with a smart under-sampling technique. The intuition behind the smart under-sampling is that most non-pulsar samples are
noise and follow the normal distribution.
We focus on removing those noise data points that perfectly fit in the normal distribution and
leave the legitimate pulsar candidates and other RFI/noise samples to be classified by the machine
learning algorithm.
Outlier detection is the identification of data points that differ significantly from other observations. In our case, the outlier detection is used to identify the data points that do not
follow the normal distribution.
In the ideal case, we hope all legitimate pulsar candidates will be identified as outliers.

\begin{equation}
D_{M}(\textbf{x}) = \sqrt{(\textbf{x}-\mu)^{T}\Sigma^{-1}(\textbf{x}-\mu)}.  \label{eq:mahalanobis}
\end{equation}

The Mahalanobis distance defined in (\ref{eq:mahalanobis})
is a measure of the distance between a point $\emph{\textbf{x}}$ and a distribution \cite{Hubert2010Minimum}.
In (\ref{eq:mahalanobis}), $\mu$ is the mean vector of the distribution, $\Sigma$ is the covariance matrix of the distribution, and the notation $T$ denotes the transpose of a matrix or a vector.
The distance is zero if $\emph{\textbf{x}}$ is at the mean $\mu$, and grows as $\emph{\textbf{x}}$ moves away from the mean along each principal component axis.

We use the Mahalanobis distance in our outlier detection.
Assume the noise points follow a normal distribution, we calculate the Mahalanobis distance
$D_{M}(\emph{\textbf{x}})$ for each candidate $\emph{\textbf{x}}$.
The mean vector $\mu$ and the covariance matrix $\Sigma$ are calculated based on the dataset.
A sample $\textbf{x}$ is considered to be an outlier if $D_{M}(\textbf{x}) \geq c$, where $c$ is a threshold.
When tuning the threshold $c$, it is required to label as many positive pulsar candidates as outliers as possible, even if there are many negative candidates labeled as outliers.
All the positive and negative samples labeled as outliers will enter Stage 2
in Fig. \ref{fig:framework} for further process.
Thus, the outlier-detection based smart under-sampling efficiently drops the negative samples
that closely follow the normal distribution.

We do not apply the smart under-sampling on the original HTRU-Medlat dataset directly, because it is difficult to find an optimal threshold $c$ with the original dataset of many attributes. Instead, we apply the smart under-sampling on the dataset of FPP diagnostic plots obtained from the HTRU-Medlat dataset. We also tested the smart under-sampling on the obtained dataset of TPP and SPP. The practical experience found that under-sampling with FPP diagnostic plots keeps all legitimate pulsar candidate and achieves a low positive-to-negative-sample ratio. After the smart under-sampling, the positive-to-negative-sample ratio reduces to nearly $1:10$, and the imbalanced data problem is alleviated.

\subsection{Stage 2: Feature Extraction}
\label{sec:stage2}

Three kinds of diagnostic plots, i.e., TPP, FPP and SPP, are used in our learning framework.
Since TPP and FPP are presented as images, and SPP is presented as a sequence, they will be
processed by different neural network models.

\begin{figure}[!t]
\centering{
\includegraphics[width=\textwidth]{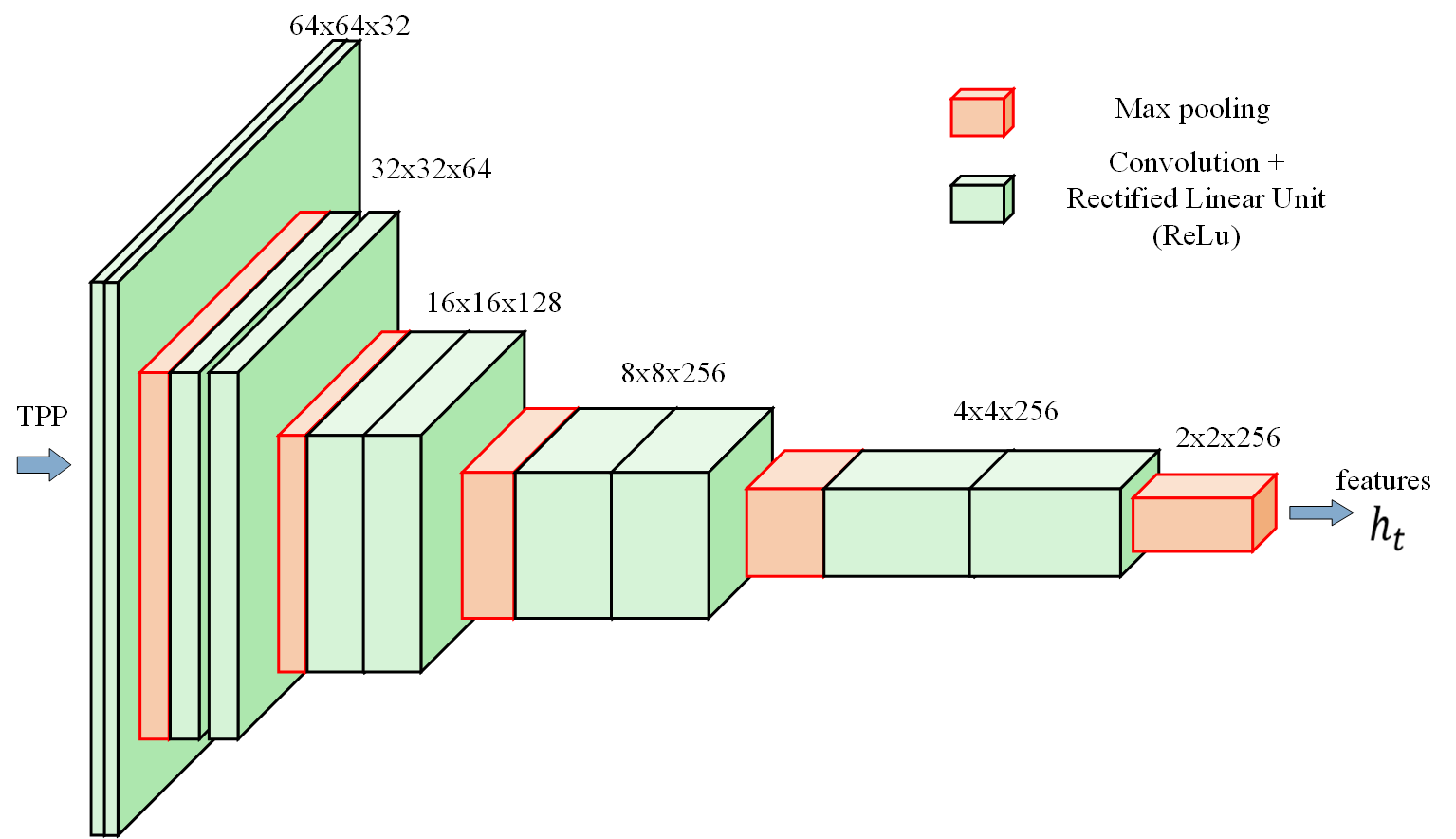}}
\caption{
The VGG-PR model for processing the TPP plots. It consists of 10 convolutional layers, 5 pooling layers and a multilayer perceptron (MLP) containing 3 fully connected layers. The output is the feature vector $h_t$. The feature vector $h_t$ will be fed into Stage 3.
\label{fig:vgg-pr}}
\end{figure}

\begin{figure}[!t]
\centering{
\includegraphics[height=7.0cm,width=\textwidth]{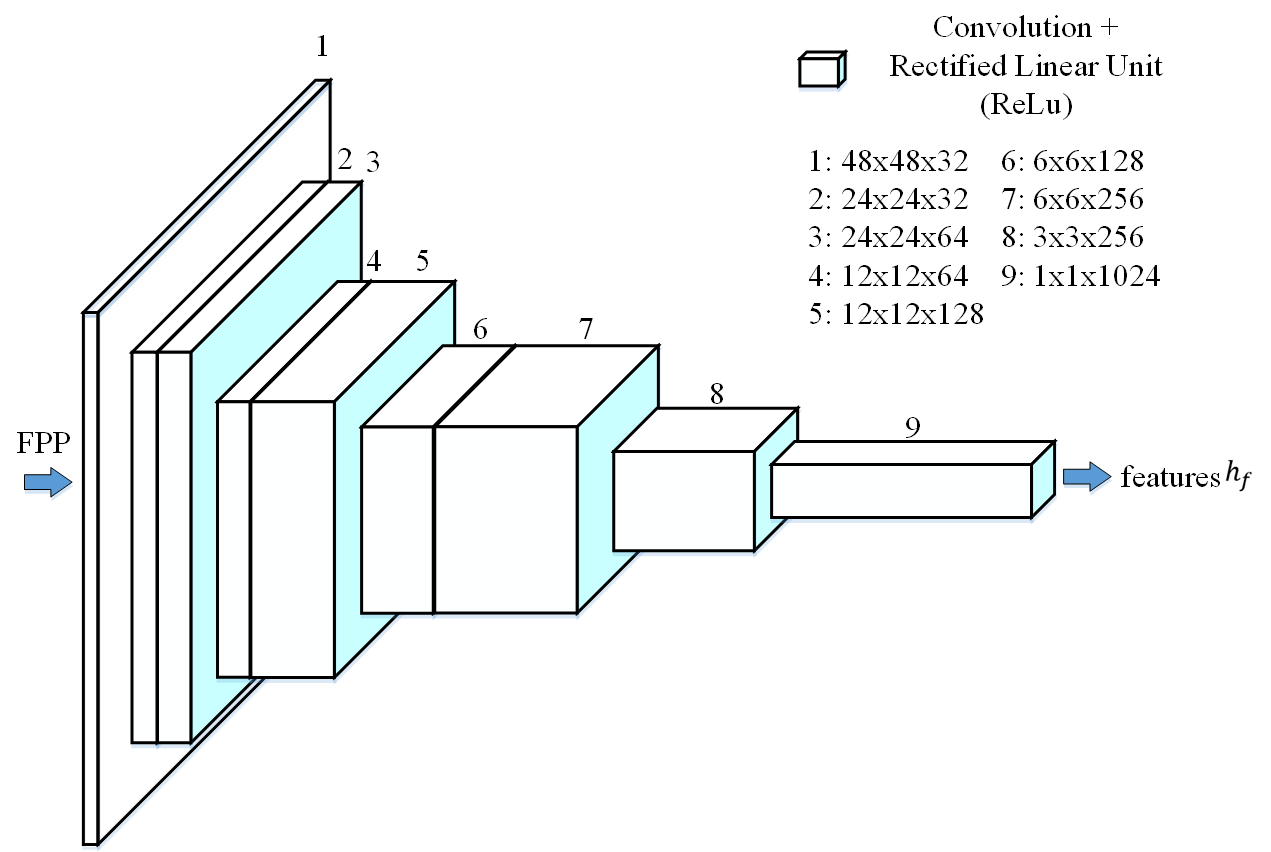}}
\caption{
The CNN-PR model for processing the FPP plots. It consists of 9 convolutional layers and a MLP  containing 3 fully connected layers. The output is the feature vector $h_f$. The feature vector $h_f$ will be fed into Stage 3.
\label{fig:cnn-pr}}
\end{figure}

Convolutional neural networks (CNN) are the most commonly applied class of deep neural networks in image processing. Deep convolutional neural network (DCNN) models that are composed of different convolutional layers and pooling layers have achieved excellent performance in image recognition.
We apply a DCNN model proposed in \cite{Simonyan2014Very} (known as VGG13) to process the TPP plots, which
is called as the VGG for pulsar recognition (VGG-PR).
As shown in Fig. \ref{fig:vgg-pr}, our VGG-PR model
consists of 10 convolutional layers, 5 pooling layers and 3 fully connected layers.
Compared with the original VGG13 model proposed in
\cite{Simonyan2014Very}, our VGG-PR model has a larger convolution
kernel and fewer channels.
For processing the FPP plots, a DCNN structure containing 9 convolutional layers and 3 fully connected layers
is used in our learning framework, which is called as the CNN for pulsar recognition (CNN-PR).
We choose different deep neural network models for processing the TPP and FPP plots,
because it helps preventing overfitting.
Note that a ResNet-like (Residual Neural Network) CNN architecture
works well in general image classification problems.
We will show in Section \ref{sec:experiment} by evaluation that our VGG-PR and CNN-PR models slightly outperform the ResNet model and justify our choice.
Both the kernels of the VGG-PR and CNN-PR models in our framework are large.
Our practice shows that DCNNs with large convolution kernels work better in pulsar candidate recognition,
even though it has been claimed in the previous work \cite{Simonyan2014Very} that multiple small convolution kernels can achieve the same performance as a large convolution kernel.

\begin{figure}[!t]
\centering{
\includegraphics[width=\textwidth]{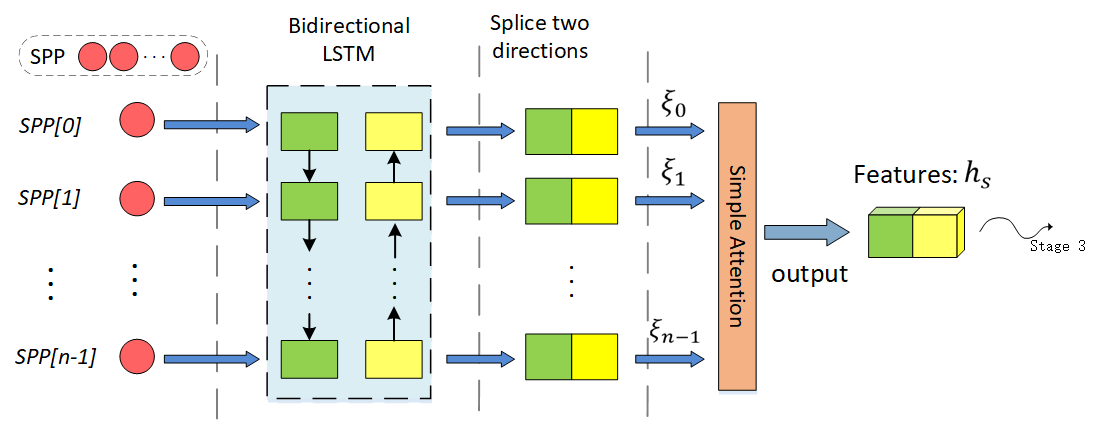}}
\caption{
The LSTM-A model for processing the SPP sequence. It consists of the bidirectional LSTM, the simple attention mechanism and a
MLP containing 3 fully connected layers. The output is the feature vector $h_s$. The feature vector $h_s$ is fed into Stage 3.
\label{fig:lstm-a}}
\end{figure}

Recurrent neural networks (RNN) are widely used in speech recognition and machine translation due to their ability of using the internal state (memory) to process sequences of inputs.
Thus an RNN model, the long short-term memory (LSTM) proposed in \cite{lakhal2018recurrent}, is used to
process the SPP sequence.
The LSTM model has three control units, i.e., the input gate, the output gate, and the forget gate,
which help the selective deletion and retention of information and make the LSTM model good at processing
long dependency sequences.
We also add a simple attention mechanism to LSTM, and thus the model is called as
the LSTM-A model \cite{long2018multimodal}.
As shown in Fig. \ref{fig:lstm-a}, the LSTM-A model
consists of the bidirectional LSTM and the simple attention mechanism.
Let the sequence of SPP be represented by an $n \times 1$ vector.
The bidirectional LSTM consists of $n$ steps, and at each Step-$i$,
the $i$th element of the SPP vector is fed into the LSTM,
where $i = 0,1,\cdots,n$.
Let $\xi_i$ denote the output vector of each Step-$i$.
The set of vectors $\xi$ will be blended by the simple attention mechanism
to form a single vector $h_s$.
More specifically, the vector $\beta$ is the learned weights through the attention mechanism.
\begin{equation}
\label{eq:kappa}
\kappa_i=\frac{e^{\beta^T \xi_i}}{\sum_{j=0}^{n}e^{\beta^T \xi_j}}, i=0, ..., n.
\end{equation}
Let $\kappa_i$ denote the weight associated with the output vector $\xi_i$,
and $\kappa_i$ can be calculated as in (\ref{eq:kappa}) .
\begin{equation}
\label{eq:h_s}
h_s=\sum_{i=0}^{n}\kappa_i \xi_i.
\end{equation}
Finally, the integrated vector by the attention mechanism, $h_s$, i.e., the feature vector of the SPP sequence is calculated as in (\ref{eq:h_s}). The feature vector $h_s$ will enter Stage 3 with the other two features
$h_t$ and $h_p$ of TPP and FPP, respectively.

\subsubsection{Improved Loss Function for the Imbalanced Dataset and Hard Samples}
Though a majority of noise data has been dropped at Stage 1, Stage 2 still
faces the problem of imbalance.
\begin{equation}
L = y log\hat{y} +(1-y)log(1-\hat{y}).
\label{eq:loss_fun}
\end{equation}
A conventional cross entropy loss function defined in (\ref{eq:loss_fun}) often leads to overfitting when the dataset is imbalanced.
In (\ref{eq:loss_fun}),
$\hat{y}$ is the probability that the label is positively predicted, and $y$ is the ground truth label.
\begin{equation}
L = \alpha y log\hat{y} +(1-\alpha)(1-y)log(1-\hat{y}).
\label{eq:alpha_loss_fun}
\end{equation}
In (\ref{eq:alpha_loss_fun}), we introduce a weighting factor $\alpha \in [0,1]$ to address the class imbalance. The weight $\alpha$ is often set as the inverse class frequency or a hyper-parameter by the cross validation in practice \cite{Lin2017Focal}.

Except for the class imbalance problem, there exist hard samples as shown in Fig. \ref{fig:hard_sample}.
\begin{equation}
L = \alpha y {(1-\hat{y})}^\gamma log\hat{y} +(1-\alpha)(1-y){\hat{y}}^\gamma log(1-\hat{y}).
\label{eq:final_loss_fun}
\end{equation}
Therefore we propose to reshape
the loss function to down-weight easy examples and thus focus on training hard negatives
as in (\ref{eq:final_loss_fun}) \cite{Lin2017Focal}.
In (\ref{eq:final_loss_fun}), the parameter $\gamma$ controls the weight of difficult-to-classify samples in the training process.

\subsection{Stage 3: Binary Pulsar Classification on Fused Features}
\label{sec:stage3}

\begin{figure}[!t]
\centering{
\includegraphics[width=\textwidth]{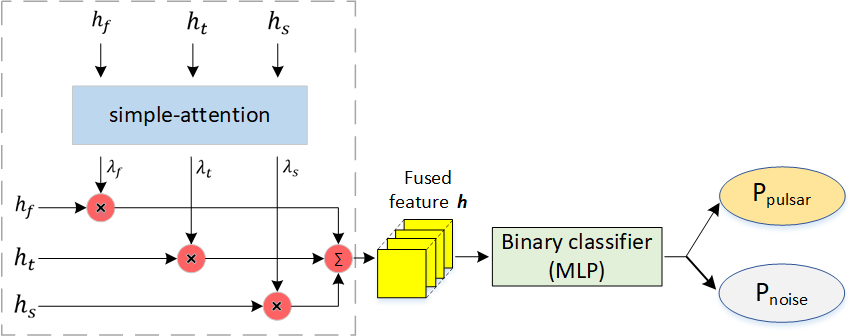}}
\caption{
Stage 3 consists of a simple attention mechanism and a binary classifier. The feature
vectors $h_t$, $h_f$ and $h_s$ obtained at Stage 2
are fed into the simple-attention layer to produce a weight vector $\omega$.
The weight vector $\omega$ is used to blend the three feature vectors $h_t$, $h_f$ and $h_s$
to form the attention vector $h$.
The attention vector $h$ is fed into the binary classifier (MLP) to produce the final sample category. The output is the probability that a sample is positive.
\label{fig:clf-a}}
\end{figure}

At Stage 2, three feature vectors are obtained from
the three diagnostic plots and need to be integrated before entering the binary classifier.
We propose to fuse features based on a simple attention mechanism proposed in \cite{long2018multimodal}.
As shown in Fig. \ref{fig:clf-a}, the feature vectors $h_t$, $h_f$ and $h_s$ are fed into
the simple-attention layer to produce a learned weight vector $\omega$.
The weight vector $\omega$ is the learned parameters in the attention mechanism.
The attention mechanism in Stage 3 is mainly used to learn the weights of the three multidimensional features, and use the learned weights to blend the three features obtained from Stage 2 to form
a single feature.
\begin{equation}
\begin{split}
\label{eq:weights}
\lambda_t = \frac{e^{\omega^T h_t}}{e^{\omega^T h_t} + e^{\omega^T h_f} + e^{\omega^T h_s}},\\
\lambda_f = \frac{e^{\omega^T h_f}}{e^{\omega^T h_t} + e^{\omega^T h_f} + e^{\omega^T h_s}},\\
\lambda_s = \frac{e^{\omega^T h_s}}{e^{\omega^T h_t} + e^{\omega^T h_f} + e^{\omega^T h_s}}.
\end{split}
\end{equation}
The weight vector $\omega$ learned from the attention mechanism is further used
to calculate the weights $\lambda_t, \lambda_f, \lambda_s$ associated with the feature vectors
$h_t$, $h_f$ and $h_s$ as in (\ref{eq:weights}), respectively.
\begin{equation}
\label{eq:h}
h = \lambda_t h_t + \lambda_f h_f + \lambda_s h_s.
\end{equation}
After obtaining the weights $\lambda_t, \lambda_f, \lambda_s$,
the integrated attention feature vector, $h$, can be calculated as in (\ref{eq:h}).
Finally, the attention vector $h$ will be fed into a binary classifier, i.e.,
the multi-layer perception (MLP) to get the final sample category.

\section{Experimental Results}
\label{sec:experiment}

In this section, we present a comprehensive experiment on the HTRU-medlat dataset,
and make a comparative analysis with the state-of-the-art results.
The experiment is conducted on a GeForce GTX 1080 graphics card with
the deep learning framework PyTorch \cite{Ketkar2017Introduction}.

\subsection{Details of Experiment Setup}

The HTRU-medlat dataset is first processed by the self-provided codes to generate TPP, FPP and SPP diagnostic plots.
Secondly, the TPP or FPP plot is resized to a $64\times64$ or $48\times48$ matrix to fit the VGG-PR or CNN-PR model, respectively. The SPP plot is represented by a $64 \times 1$ vector.
Finally, all the elements of the TPP/FPP/SPP matrices or vectors are normalized to remove the absolute scale of the plots.
\begin{align}
\label{eq:normalize}
v_i = \frac{v_i - \mu}{\sigma}.
\end{align}
More specifically, for a vector $v$, each element $v_i$ is re-calculated as in (\ref{eq:normalize}),
where $\mu$ and $\sigma$ are the mean and standard deviation of the elements in vector $v$.
For a matrix, normalization is performed by row vectors, which removes
instrumental variations but remains the variance in signal \cite{osti_22348078}.

At Stage 1 of data cleaning, we need to tune the threshold $c$ to keep more than $99\%$ of positive candidates.
As we discussed before, we do the smart under-sampling only using the FPP diagnostic plots. We flatten a $48\times48$ FPP matrix to a $2304 \times 1$ vector.
The vector is further reduced to a $32 \times 1$ vector by the technique of the principal component analysis (PCA) \cite{Abdi2010Principal}.
All the samples are sorted in descending order according to the Mahalanobis distances, and the threshold $c$ can be determined by keeping more than $99\%$ of positive samples in the dataset.
\begin{table}
\begin{center}
\caption[]{Percentage of Remaining Positive Samples with Different $c$}\label{tab:stage1}
 \begin{tabular}{cccc}
  \hline\noalign{\smallskip}
c& recall of positive& recall of negative& filtering rate                   \\
  \hline\noalign{\smallskip}
200 &48.2\% &1.4\% &2.05\% \\
100 &72.2\% &2.7\% &3.65\% \\
50  &97.8\% &6.5\% &7.82\% \\
\textbf{43}  &\textbf{99.8\%} &\textbf{13.8\%} &\textbf{15.00\%} \\
  \noalign{\smallskip}\hline
\end{tabular}
\end{center}
\end{table}
In Table \ref{tab:stage1}, we show the percentage of positive samples that remain in the dataset with different $c$.
We choose $c=43$ and remain about $99.8\%$ positive samples and $13.8\%$ negative samples after cleaning.
The filtering rate is defined as the number of remaining samples over the number of original samples.
At Stage 2, the VGG-PR, CNN-PR and LSTM models follow three fully connected layers with hidden size 1024, 256, and 256, respectively. The activation function in fully connected layers is PReLU \cite{He2015Delving}.
The obtained features are represented by three $1024\times1$ vectors, respectively.
At Stage 3, the model MLP consists of three fully connected layers with hidden size 1024, 256, and 256, respectively.


\subsection{Evaluation Criteria}

\begin{equation}
Recall=  \frac{TP}{TP+FN}
\label{eq:recall}
\end{equation}

\begin{equation}
Precision=  \frac{TP}{TP+FP}  \label{eq:precision}
\end{equation}

\begin{equation}
F1=  2 \times \frac{Recall \times Precision}{Recall+Precision}
\label{eq:F1}
\end{equation}

We use $F1$ score to evaluate our approach considering the imbalance between positive samples and negative samples in HTRU medlat dataset \cite{Yao2017Pulsar}.
The $F1$ score is defined as in (\ref{eq:recall})-(\ref{eq:F1}),
where $TP$ is the number of positive samples that are predicted as positive samples; $FN$ is the number of positive samples that are predicted as negative samples; $FP$ is the number of negative samples that are predicted as positive samples.
We only evaluate the $F1$ score for positive samples, because
only efficient recognition of positive samples is meaningful in pulsar searching.

\begin{equation}
TPR=  \frac{TP}{TP+FN}  \label{eq:TPR}
\end{equation}

\begin{equation}
FPR=  \frac{FP}{TN+FP}  \label{eq:FPR}
\end{equation}

We also define the True Positive Rate (TPR) and False Positive Rate (FPR) as in (\ref{eq:TPR})
and (\ref{eq:FPR}), respectively. The TPR and FPR are used to plot the Receiver Operating Characteristic (ROC) curve, which can effectively evaluate the trained models of the unbalanced dataset \cite{carter2016roc}. The area under the ROC curve (AUC) is a commonly used
measure of the model's capability of distinguishing between classes. A higher AUC generally stands a better model at distinguishing between classes.

\subsection{Detailed Results}

\begin{table}
\begin{center}
\caption[]{Experimental Results Compared with the State-of-the-art Works.}
\label{tab:compare_results}
 \begin{tabular}{lcccc}
  \hline\noalign{\smallskip}
        &$F1$-score &$Recall$ &$Precision$
     &time (minute)    \\
  \hline\noalign{\smallskip}
\cite{Guo2017Pulsar}  &0.96 &0.96 &0.96 &$\approx10^3$\\
Our approach  &0.95 &0.94 &0.96 &$\approx10^2$\\
  \hline\noalign{\smallskip}
\cite{Zhang2019Pulsar} (average results)  &0.92 &0.94 &0.91 &$\approx10^2$ \\
Our approach (average results) &0.95 &0.95 &0.96 &$\approx10^2$\\
  \noalign{\smallskip}\hline
\end{tabular}
\end{center}
\end{table}

In Table \ref{tab:compare_results}, we compare our approach with the state-of-the-art works in \cite{Guo2017Pulsar} and \cite{Zhang2019Pulsar}.
When we compare with the work in \cite{Guo2017Pulsar},
the HTRU medlat dataset after under-sampling is split in to three parts with a ratio of $4:3:3$, for training, validation and testing, respectively, and the results of a single test are reported.
When we compare with the work in \cite{Zhang2019Pulsar},
we follow a 10-foldcross validation that is used in \cite{Zhang2019Pulsar}.
In the 10-foldcross validation, the HTRU-medlat dataset after under-sampling is randomly split into 10 subsets, and 9 subsets are used as the train sets and the 10th subset is used as
the test set. The procedure is repeated randomly for 10 times and the average
performance is reported in Table \ref{tab:compare_results}.
As shown in Table \ref{tab:compare_results},
the performance of our approach is comparable with that of \cite{Guo2017Pulsar} in terms of $F1$ score,
$Recall$ and $Precision$,
and our approach improves the training time by $10X$.
Compared with the work in \cite{Zhang2019Pulsar},
our approach has a similar training time and outperforms it in all other criteria.

\begin{table}[t]
\caption{Compare with the ResNet34 model.}
\label{tab:resnet}
\begin{center}
\begin{tabular}{ccccccc}
\hline
& \multicolumn{3}{c}{VGG-PR/CNN-PR} & \multicolumn{3}{c}{ResNet34} \\
\hline
& $F1$ score & $Recall$ & $Precision$  & $F1$ score & $Recall$ & $Precision$ \\
\hline
TPP & 0.95 & 0.96 & 0.94 & 0.94 & 0.94 & 0.93\\
\hline
FPP & 0.92 & 0.93 & 0.90 & 0.91 & 0.91 & 0.91 \\
\hline
\end{tabular}
\end{center}
\end{table}

We justify the choice of VGG-PR model and CNN-PR model for processing the TPP and FPP plots, respectively.
In the experiments, we replace the VGG-PR or the CNN-PR model by a ResNet34 model proposed in \cite{he2015deep}.
Table \ref{tab:resnet} shows the performance of the ResNet34 model, which is not good enough compared with
that of VGG-PR/CNN-PR.

\begin{table}
\begin{center}
\caption[]{Classification Based on the Fused Features vs Classification Based on Each Individual Feature.}
\label{tab:feature_fuse}
 \begin{tabular}{lccc}
  \hline\noalign{\smallskip}
Feature &$F1$ score & $Recall$ & $Precision$                    \\
  \hline\noalign{\smallskip}
FPP  &0.92 &0.94 &0.90\\
TPP  &0.95 &0.94 &0.96\\
SPP  &0.89 &0.94 &0.85\\
Fused Features  &\textbf{0.96} &\textbf{0.95} &\textbf{0.97}\\
  \noalign{\smallskip}\hline
\end{tabular}
\end{center}
\end{table}

\begin{figure}[!t]
\centering{
\includegraphics[width=9.0cm]{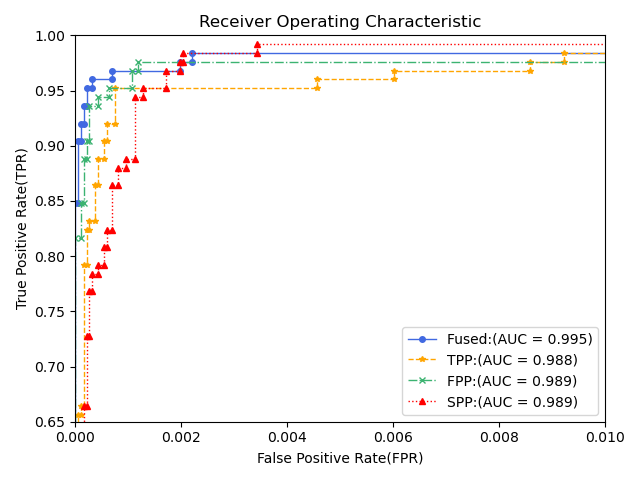}}
\caption{The ROC Curves of Classification Based on the Fused Features or Each Individual Feature.
\label{fig:roc}}
\end{figure}

\begin{table}
\begin{center}
\caption[]{Comparison of different fusion strategies.}
\label{tab:compare_fusion}
 \begin{tabular}{lccc}
  \hline\noalign{\smallskip}
 &$F1$ score & $Recall$ & $Precision$                    \\
  \hline\noalign{\smallskip}
concatenation  &0.95 &0.96 &0.94\\
weighted summation &0.95 &0.96 &0.94\\
attention-based &0.96 &0.97 &0.95\\
  \noalign{\smallskip}\hline
\end{tabular}
\end{center}
\end{table}

Next, we evaluate the attention-based feature fusion strategy.
We do the classification based on each individual feature of TPP, FPP or SPP instead of the fused features obtained at Stage 3. we also follow the 10-fold cross validation.
In Table \ref{tab:feature_fuse}, we compare the performance of classification based on the fused features with the classification based on each individual feature.
It is obvious that the feature fusion based approach improves all evaluation criteria including $Recall$, $Precision$ and $F1$ score greatly.
In Fig. \ref{fig:roc}, we present the ROC curves of the classification.
Note that the TPR and FPR are calculated based on the original HTRU-medlat dataset.
It shows that the classification with the fused features obtains a high TPR while maintains a low FPR and achieves the highest AUC.
We further compare the attention-based feature fusion strategy with the concatenation and weighted summation fusion strategies proposed in \cite{zhang2020}. The concatenation strategy simply concatenates all the three
features, and the weighted summation fusion strategy uses the weighted sum of the three features.
We simply replace our attention-based feature fusion by the two candidate strategies in the experiments.
Table \ref{tab:compare_fusion} shows that the attention-based strategy achieves the best performance compared
with the other two strategies.
Overall, the proposed feature fusion benefits the pulsar candidate recognition.

\begin{table}
\begin{center}
\caption[]{The Improved Loss Function vs the Original Focal Loss Function.}
\label{tab:compare_loss}
 \begin{tabular}{lccc}
  \hline\noalign{\smallskip}
 &$F1$ score & $Recall$ & $Precision$                    \\
  \hline\noalign{\smallskip}
Improved Loss Function  &0.95 &0.94 &0.96\\
Original Loss Function  &0.95 &0.93 &0.97\\
  \noalign{\smallskip}\hline
\end{tabular}
\end{center}
\end{table}

We compare the performance of the improved loss function with the original focal loss function
in Table \ref{tab:compare_loss}.
The results show that both the loss functions obtain the same $F1$ score of $0.95$. Nevertheless, the improved loss function can achieve a more balanced combination of $Recall$ and $Precision$.

\begin{figure}[!t]
\centering{
\includegraphics[width=9.0cm]{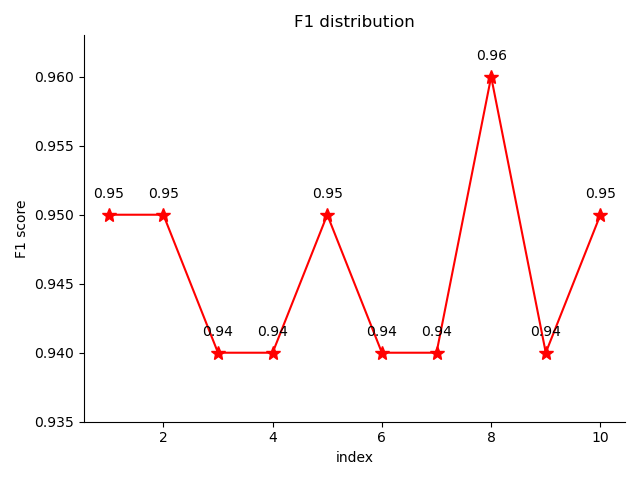}}
\caption{The performance results of $10$ random experiments.
\label{fig:average_result}}
\end{figure}

Finally, we evaluate the robustness of our approach by randomly splitting the dataset for $10$ times. We report $F1$ score of each individual
test and the average performance of the tests in Fig. \ref{fig:average_result}. It shows
that the average obtained $F1$ score
is about $95\%$, and all random tests achieve the $F1$ scores around
$94\%-96\%$. 

\section{Conclusion}
\label{sec:conclusion}

We propose a deep learning framework consisting of the smart under-sampling, early feature fusion and the improved loss function for pulsar recognition to cope the class imbalance problem.
Considering the F1-score of positive samples and the training time, our framework can get a competitive result and speed up the training time by $10X$ compared with the state-of-the-art works.
We conclude our unique contribution. First, we propose a strategy for under-sampling based on the Mahalanobis distance, which drops most of the negative samples. Secondly, we use the simple attention mechanism to fuse features extracted by the artificial neural networks. Thirdly, we improve the cross entropy loss function for the imbalanced class and hard negative sample issues.
To our best knowledge, it is the first time that the improved loss function is used in pulsar recognition.
All the techniques we proposed are not only helpful in pulsar identification, but also can be used to solve other extremely unbalanced problems.


\normalem
\begin{acknowledgements}
This work was supported in part by Science and Technology Commission of Shanghai Municipality, China (Grant No. 19ZR1463900) and National Key R$\&$D Programme of China 2018YFA0404603.
\end{acknowledgements}

\bibliographystyle{raa}
\bibliography{pulsar}

\label{lastpage}

\end{document}